# Astrophotonic spectroscopy: defining the potential advantage

Jeremy Allington-Smith

Centre for Advanced Instrumentation, Physics Dept, Durham University, South Rd, Durham DH1 3LE

and Joss Bland-Hawthorn

School of Physics, University of Sydney, NSW 2206, Australia

**Abstract**

A photonic spectrograph can be much smaller than a conventional spectrograph with the same resolving power. Individual devices can be integrated with optical fibres to improve the multiplex gain in astronomical spectroscopy. Although experimental devices have been tested, the parameter space where integrated photonic spectrographs give a significant advantage over traditional methods has not been defined. This paper gives an overview of the theory with verification by direct simulation using Fresnel propagation and quantifies the benefit for representative spectroscopic capabilities. We thereby confirm the advantage of photonic spectrographs, especially to the next generation of Extremely Large Telescopes, and therefore conclude that these devices may be important for the future development of astronomical instrumentation.

## 1 Introduction

In principle, it is possible to obtain a given resolving power with a much reduced beam size by using waveguide arrays instead of bulk dispersers. This could allow optical fibres to be directly integrated with the spectrographs (Bland-Hawthorn and Horton 2006; BH) to the benefit of highly-multiplexed spectroscopy in multiobject or diverse-field configurations (Murray and Allington-Smith 2009, Poppett, Allington-Smith and Murray 2009). The devices can also be combined with other photonic systems to filter out spectrally-structured telluric background radiation (Bland-Hawthorn et al. 2004). The photonic approach also allows simplification of the device by eliminating with the need for bulk optics

These Integrated Photonic Spectrographs (IPS) may be amenable to mass-production using photonic fabrication techniques such as Ultrafast Laser Inscription (ULI: Thomson, Kar and Allington-Smith 2009 – hereafter TKA) or Hybrid Sol-Gel devices (Ghasempur et al. 2009) and so facilitate highly-multiplexed spectroscopy on very large scales and defeat the harsh scaling laws which apply to Extremely Large Telescopes (ELTs; Allington-Smith 2007). One possible geometry is shown in Figure 1.

Why the normal theoretical derivation for the resolving power of a spectrograph – which imply large dispersers (100-300mm) - can be circumvented has not been discussed in the literature (except briefly by TKA). This has prevented quantification of the potential advantage when considering instrumentation for ELTs, hampering long-term planning for these vital research tools.

In this paper, the potential advantage to astronomy is investigated theoretically (§2) and by direct calculation using Fresnel propagation for generic IPS layouts (§3). Finally, the design parameters for astronomically-useful spectrographs are derived (§4) as an aid to further work.

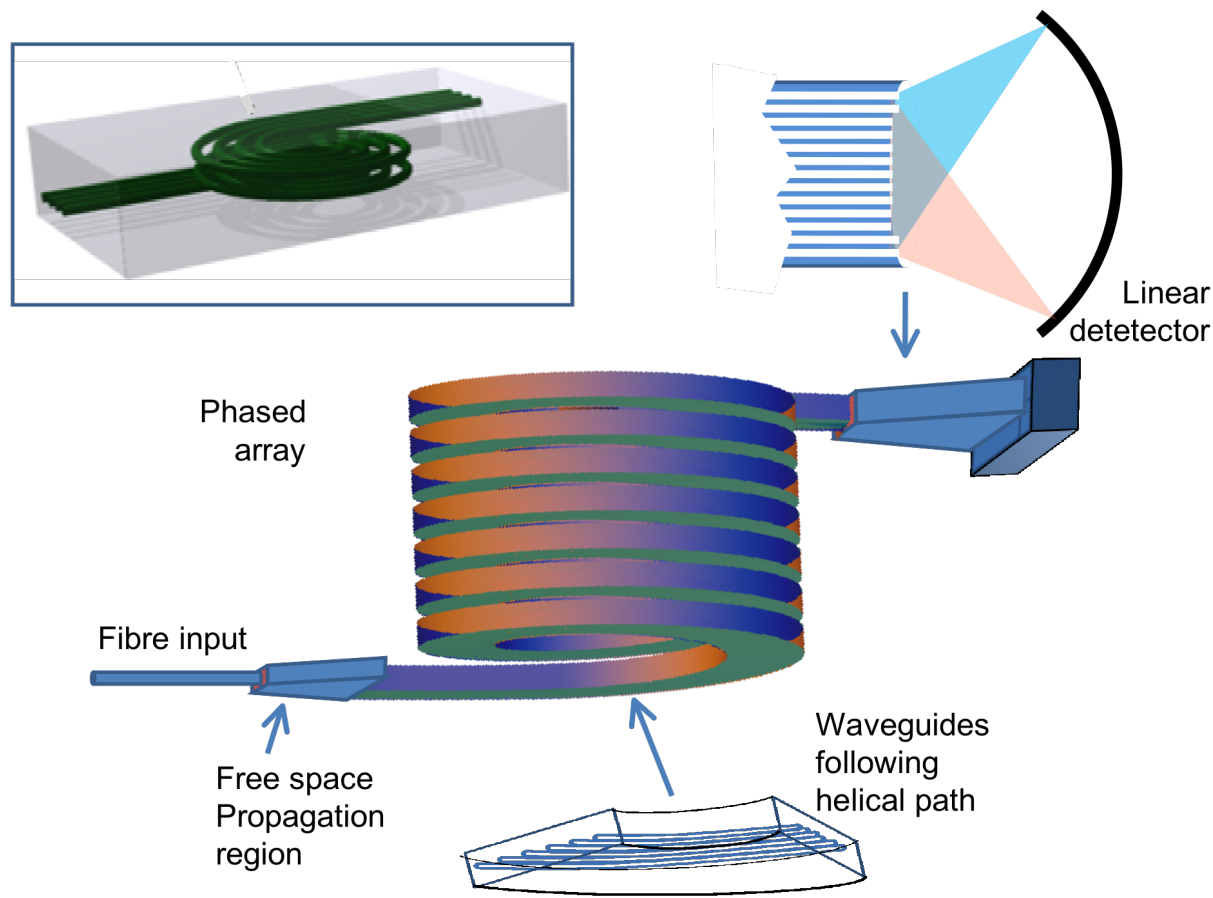

**Figure 1: One possible IPS layout in which the path delays between waveguides is generated by winding the array helically. The inset at top-left represents a possible ULI implementation (TKA).**

## 2 Theory

Conventional slit-limited spectroscopy gives the resolving power in terms of the grating blaze angle, which is identical to the grating tilt at blaze in a Littrow configuration, as

$$R \equiv \frac{\lambda}{\delta\lambda} = \frac{2D\tan\gamma}{\chi D_T} \qquad (1)$$

where $\delta\lambda$ is the wavelength resolution (e.g. Lee and Allington-Smith 2000; Bland-Hawthorn and Cecil 1997). Since this involves only the diameters of the telescope aperture and collimated beam ($D_T$ and $D$ respectively) and the angular slitwidth, $\chi$, it is generally assumed that this imposes a fixed upper limit on the obtainable resolving power.

However this equation is only true if the disperser is a conventional diffraction grating fashioned as a single slab of optical material. As stated by BH and TKA, the limitations implied by (1) may be circumvented through the use of micro-waveguide technology to construct a waveguide network to act as a grating by introducing arbitrary phase delays between waveguides. Since $R$ is simply the ratio of the maximum path difference introduced by the disperser to that caused by the finite width of the slit, the required path differences may be produced by varying the length of adjacent waveguides instead of tilting a bulk disperser such as a diffraction grating. Furthermore, the use of waveguides on small scales allows traditional optical elements to be eliminated with consequent savings in size, weight and cost.

Throughout this paper IPS will be compared with a Simplified Traditional Spectrograph (STS) operating in the Littrow configuration via an inclined monolithic disperser: either a surface-relief reflection or transmission grating or an equivalent volume phase holographic grating. The general disperser geometry is indicated in Figure 2 but only the Littrow configuration for which, $\Psi = \alpha + \beta = 0$ is considered. The disperser is coupled to a collimator and camera (not shown) whose details are irrelevant here. The figure also illustrates the problem of groove shadowing which reduces throughput when $\psi$ is large.

The grating equation giving the angle of diffracted light, $\theta$, measured with respect to the disperser normal, in order $m$ for a disperser consisting of scattering centres equally spaced with linear density $\rho$, is

$$\sin\theta + \sin\theta_I = (m\lambda + q)\rho \qquad (2)$$

where $q$ is an extra path difference between successive centres and $\theta_I$ is the angle of incidence. The angles is that they have the same sign if the rays are on the same side of the axis in question.

For a conventional spectrograph, $q = 0$, so the diffracted angle in order $m$ is given by $\sin\theta = m\lambda\rho - \sin\theta_I$. The angular dispersion is obtained by differentiating the grating equation with respect to the exit angle,

$$\Delta \equiv \frac{d\lambda}{d\theta} = \frac{\cos\theta}{m\rho} \qquad (3)$$

which is independent of the extra path difference. Thus to obtain a desired dispersion, $\Delta_P$, the required diffraction order is

$$m_P = \frac{\cos\theta_P}{\rho\Delta_P} \qquad (4)$$

For the light to exit at angle $\theta_P$, the path difference at wavelength $\lambda_0$ must be

$$q = \frac{\sin\theta_P + \sin\theta_I}{\rho} - m_P\lambda_0 = \frac{1}{\rho}\left(\sin\theta_P + \sin\theta_I - \frac{\lambda_0}{\Delta_P}\cos\theta_P\right) \qquad (5)$$

from (2) and substituting for the spectral order from (4).

We wish to minimise this angle so that it will be near the maximum in the blaze function for a waveguide with its axis normal to the plane of the disperser. Thus we can obtain higher dispersion by adopting a high diffraction order and use the extra path difference to adapt the exit angle to satisfy the blaze condition.

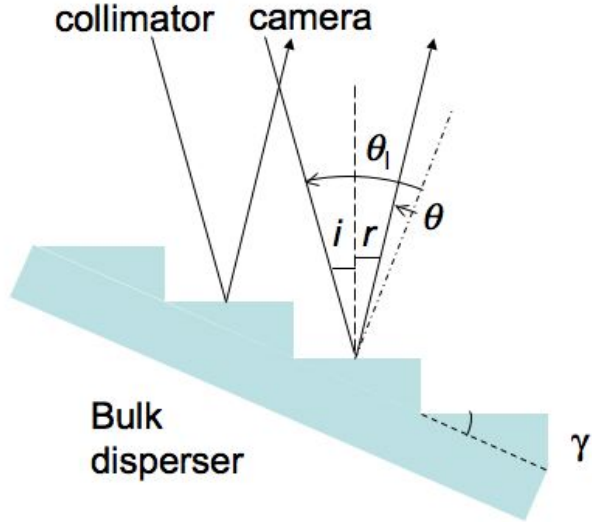

**Figure 2: Illustration of the blaze condition for a conventional surface-relief reflection grating. The dashed and dash-dot line are normal to the facet and plane of the disperser respectively.**

For the STS this occurs when the phase difference between the centre and edge of one facet,

$$\phi = \frac{\pi \cos\gamma}{\rho\lambda}(\sin i - \sin r) = 0 \qquad (6)$$

to maximise the blaze function. For a reflection grating (for which $\theta < 0$) this occurs when $i = r$ so that, from (2), $2\sin\gamma = m_Z\lambda\rho$ in order $m_Z$.

For a photonic disperser using a phased array the blaze condition is $\theta_I = \theta_P = 0$ since $\theta_I$ is defined by the symmetry axis of the waveguide exit assumed to be normal to the plane containing the local waveguide exits. If this condition is satisfied for wavelength $\lambda_0$, the phase to be added at wavelength $\lambda$ is

$$\phi(\lambda) = \frac{2\pi q}{\lambda} = -2\pi \frac{\lambda_0}{\lambda} \frac{1}{\rho\Delta_P} \qquad (7)$$

Thus the dispersion may be increased by an arbitrary factor determined by the size of the path length, $q$.

In effect an IPS operates in a arbitrarily high spectral order but without the disadvantage of the diffracted light exiting at very large angles to satisfy the blaze condition. Thus the IPS is blazed for low angles, close to the optical axis despite working in a high spectral order. For an STS the high angle requires the monolithic disperser to be tilted at a large angle and so takes up a lot of space. The same could be done with a photonic disperser of much smaller size by compacting (e.g. coiling up) the waveguides without any need to tilt the grating. Furthermore, any limit on the grating angle this does not necessarily apply to a photonic disperser. Thus there is the potential for a more compact spectrograph.

The spectral resolution is determined by the wavelength range covered by the image of a monochromatic point source at the slit. For a slit-limited spectrograph (i.e. with a slit sized to match an image larger than the diffraction limit of the telescope), this is

$$\delta\lambda = \frac{d\lambda}{d\theta}\frac{d\theta}{dx}\delta x' = \Delta\frac{\delta x'}{f'} = \Delta\frac{\delta x}{f} = \frac{s\Delta}{FD} \qquad (8)$$

where, respectively, $f$ and $f'$ are the focal lengths of the collimator and camera and $\delta x$ and $\delta x'$ are the sizes of the slit and its image at the detector (linked by the conservation of Etendue), $s = \delta x$ is the physical slitwdth, $F$ is the collimator focal ratio and $D$ is the diameter of the collimated beam. Therefore for the same slitwidth and collimator speed, the beam diameter can be reduced by the ratios of the dispersion, i.e. $D_P/D_Z = \Delta_P/\Delta_Z$ where

$$\Delta_Z = \frac{\cos\gamma}{m_Z\rho} = \frac{\lambda_0}{2\tan\gamma} \qquad (9)$$

is the dispersion for an STS obtained from (3) at the blaze condition $2\sin\gamma = m_Z\lambda\rho$. The equivalent for an IPS with the same exit angle is obtained from eqn (4) and (2) with $\theta_P = 0$

$$\Delta_P = -\frac{\lambda_0}{q\rho} \qquad (10)$$

which can be arbitrarily small. Therefore the photonic size advantage is

$$\frac{D_P}{D_Z} = \frac{\Delta_P}{\Delta_Z} = \frac{2\tan\gamma}{q\rho} \qquad (11)$$

which vanishes as $q \to \infty$ since for a practical spectrograph $\gamma < \gamma_{max}$ where, for current large echelle gratings, $\gamma_{max} = \arctan(4)$.

The size advantage refers to the beam size whereas the size of a spectrograph is also determined by the size of the input and output fields, at the slit and detector. However this is not a major consideration if the spectrograph is to be integrated with individual fibres since the fibre field is then very small (typically 0.1mm) and the detector is one-dimensional rather than two-dimensional. Since this type of integration is one of the justifications for the use of photonic principles, (11) is an appropriate metric to describe the size advantage.

Another important parameter is the number of scattering centres, $N$, which defines the number of combining beams. For an IPS, this is defined by the waveguides in a photonic disperser or rulings for an STS. For a diffraction-limited spectrograph, this is related to resolving power via $R = mN$. More generally, $N$, is the Finesse, which is equal to the number of combining beams in a perfect spectrograph in the diffraction limit but has other interpretations for interferometric spectrographs such as Fabry-Perot interferometers.

However, in most astronomical applications the slitwidth significantly exceeds the diffraction limit of the telescope, so the number of scattering centres can be reduced to

$$N_{min} = \frac{R\lambda}{m\chi D_T} < \frac{R}{m} \qquad (12)$$

However the actual number required may be higher than this. For example, when tested, a commercial device adapted for astronomy with 400 waveguides (Cvetojevic et al. 2009) gave $R \approx 2000$ with $m = 26$ although the calculated $N_{min} \approx 80$. This system is diffraction limited already so $N/N_{min} \approx 5$ represents a departure from the simple theory which can be explained by non-uniform illumination of the beams, cross-talk between waveguides or defects in construction. This is reflected in the fabricators' rule-of-thumb that the number of waveguides needs to exceed the theoretical requirement by a factor 4.

## 3   Direct verification via Fresnel propagation

Although the theoretical expectation is quite clear, it is based on Fraunhoffer diffraction theory. This is commonly used for astronomical optics but needs to be verified on the much smaller scales of photonic devices where the approximations of Fraunhoffer theory may be invalid. Therefore, to verify that a photonic disperser can produce arbitrarily high dispersion while blazed at low angle, the operation of simplified photonic disperser has been simulated by exact calculation of Fresnel propagation making only the simplest assumptions about the initial electric field and the geometry.

The electric field is specified as a scalar complex number at the disperser ignoring polarisation effects. Fresnel propagation through free space then gives the electric field in the plane of a detector, which yields the light intensity.

It is useful to first consider how to model the process with a simple linear geometry in the *x-z* plane where *z* defines the distance along the normal to the disperser towards a detector defined with coordinate *x* (**Figure 3**). The plane of the disperser is parallel to this with coordinates ($x'$, $z = z_0$).

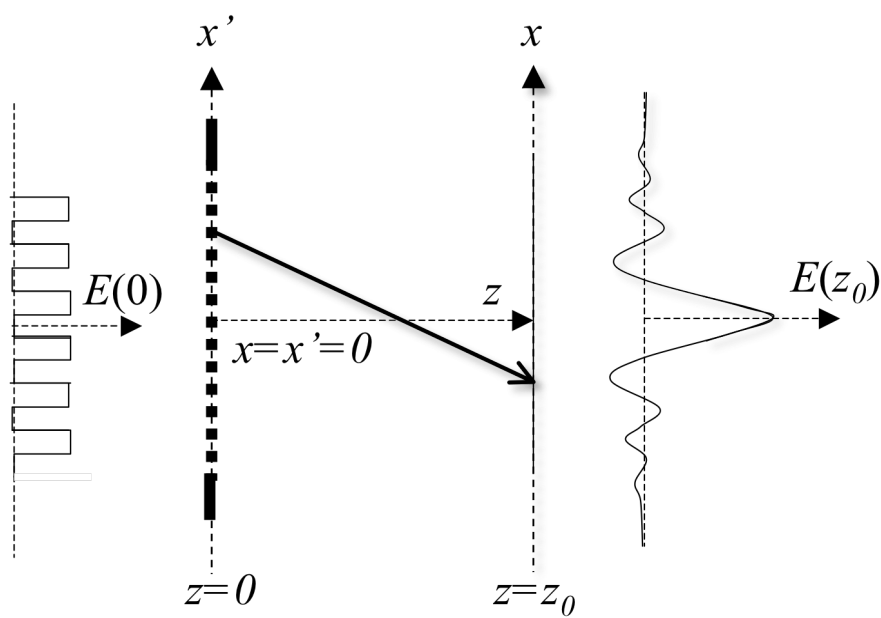

**Figure 3: Geometry used in Fresnel propagation calculation illustrating the real part of the input and output electric fields. $z = 0$ is the plane of the disperser where the electric field in the waveguides is $E(0)$. Light propagates freely to the image plane at $z = z_0$ where the electric field is $E(z_0)$. $x$ and $x'$ are coordinates parallel to the dispersion at the image and disperser planes respectively. For clarity, we use planar geometry in the figure. See the text for details of how this is modified in practical cases.**

To obtain the electric field strength on the detector surface at $(x, z = z_0)$ we apply the Fresnel propagation operator

$$F(r,\theta,k) = \exp[-ikr]\cos\theta \qquad (13)$$

via the transformation $r^2 = (x-x')^2 + z_0^2$ and $\theta = \arctan[(x-x')/z_0]$, so that

$$E(x,z) = \int E(x',0)F(x - x',z,k)dx' \qquad (14)$$

The scattering centres are modelled as apertures of extent $b < a = 1/\rho$ with electric field strength unity and phase either zero or as given by equation 7.

In practice, the geometry of **Figure 3** must be modified to focus the spectrum on the detector surface unless the image-disperser separation is very large. Figure 4 shows some possible configurations compared to the planar geometry. These use combinations of path differences (indicated by hollow lines) or the curvature of the disperser to obtain the same effect. In these cases the expressions for $r$ and $\theta$ in (13) are modified to reflect the curved geometry. Further configurations exist if the detector is not constrained to be planar. Note that the 1-D nature of the detector makes it relatively simple to curve.

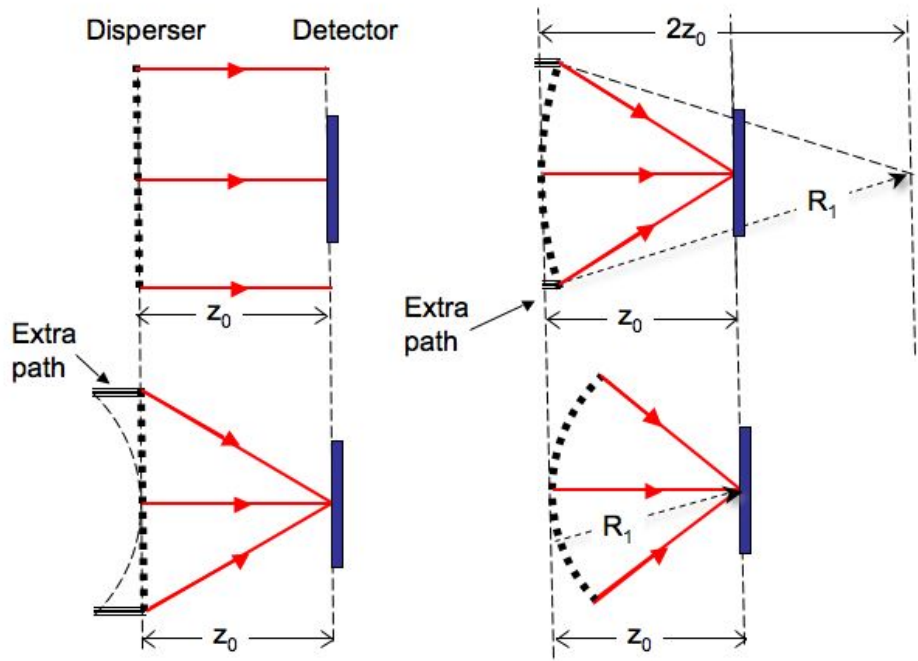

**Figure 4: Modified geometries to produce images without additional optics. Top-left - the simplified geometry illustrated in Fig. 3. The three other cases show configurations which produce focussed images of the spectrum on the detector surface (here constrained to be flat) using curvature of the disperser or the introduction of extra path differences by modifying the waveguides. Top-Right – geometry used for the predictions; Bottom – alternative layouts. The electric field in the space to the left of the dashed line is uniform. Where needed, extra path-differences are introduced into the waveguides as indicated by the hollow lines.**

Figure 5 shows the resulting intensity on the detector surface. By measuring the centroid of the profile of features in first order ($m_Z = 1$) with and without the phase shift, we confirm that the dispersion is increased by a factor $m_P/m_Z$. Figure 5 gives an example. In this case, two curves are shown for different wavelengths where the difference is scaled by $m_Z = 1$ and $m_P = 100$ so that the expected angular displacement is the same and the peak of the profile occurs at the expected location of the first order light from a conventional grating, $m_Z = 1$. It can be seen that this is indeed the case. For the photonic spectrograph the obtained linear dispersion is $6.63 \pm 0.1 \times 10^{-6}$ compared with the prediction of $6.67 \times 10^{-6}$

The simulation also confirms that the resolving power is given by the product of the dispersion order and the number of scattering centres. In the example shown the number of scattering centres is 100. This yields R=115 for the traditional spectrograph and $R = 1150$ for the photonic spectrograph with an estimated 5% fitting error based on estimating $\delta\lambda$ as $2.355\sigma$ where $\sigma$ is obtained from a Gaussian fit. The difference from the expected values of 100 and 10000, respectively, is probably due to the fact that the line profile deviates from the Gaussian function used in the fitting.

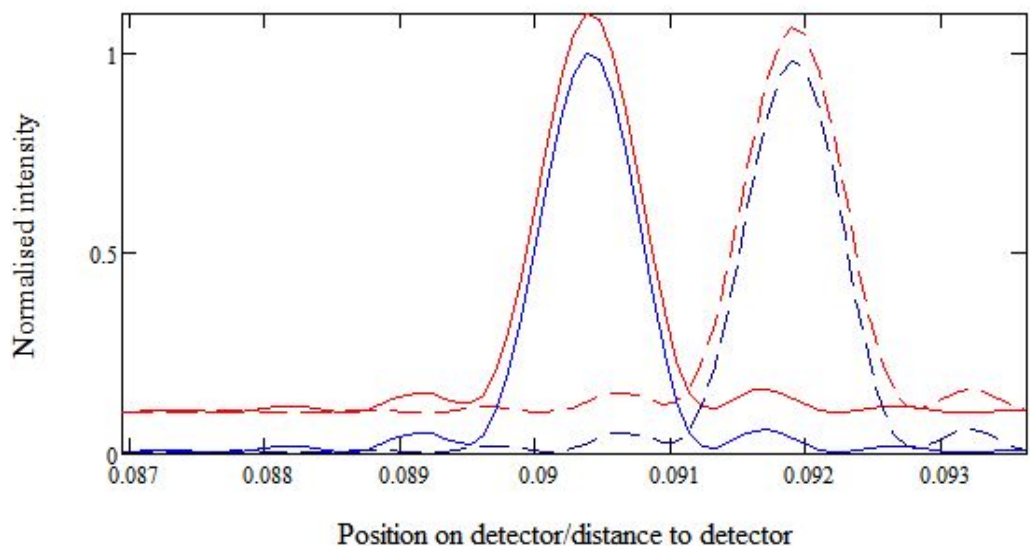

**Figure 5: Verification of dispersion by direct Fresnel propagation simulation. The red curves are predictions for the STS and the blue curves are for the IPS working in effective order 100. The solid lines are for wavelength 600nm and the dashed curves are for 610nm for the traditional spectrograph and 600.1nm for the photonic spectrograph. The position on the detector is the same indicating that the dispersion differs by the expected factor of 100. The spectral resolution also differs by this ratio since the spatial width of the line profile is the same in each case. The unphased predictions are offset by 0.1 in the abscissa. The intensity is normalised to unity for the unphased 600nm prediction.**

## 4  Practical examples

The utility of photonic spectrographs is now examined by considering representative applications for 8 and 40m telescopes as follows.

A. $R = 50000$ at $\lambda_0 = 50$nm using fibres in natural seeing, ideal for Galactic Archaeology

B. $R = 5000$ at $\lambda_0 = 1700$nm with a moderately narrow slit, ideal for SNR observations of high-redshift objects via optimal rejection of night-sky features

C. $R = 1000$ at $\lambda_0 = 500$nm with a narrow slit, for traditional slit spectroscopy in the visible with AO

D. $R = 100$ at $\lambda_0 = 700$nm for large-scale structure mapping in natural or partly-corrected seeing.

The chief instrumental parameters for a conventional and photonic spectrograph are calculated for three sets of assumptions for the traditional spectrograph. In all cases the pixel size is set to 15μm and the camera focal ratio, $F'$, adjusted so that the sampling aperture is spanned by 4 detector pixels. Results are presented for 8m and 40m telescopes in Table 1 and Table 2 respectively.

For the conventional spectrograph, there are three options relating to the disperser tilt.

(1) With beam sizes typical of current traditional spectrographs on 8-10m telescopes in the range 50-300mm.

(2) With the beam size reduced until $\tan\gamma = 1$. This represents the maximum practical blaze angle for use in first order to maximise both efficiency and free spectral range

(3) As (2) but for $\tan\gamma = 4$. This represents the maximum feasible blaze angle due to grating ghosts and polarisation effects irrespective of efficiency and free spectral range. This limit is actually imposed by the maximum ruled area that fabricators can currently produce (a 2 x 1 mosaic of ~ 440 x 480mm). Although this issue is irrelevant for this discussion, there are other problems caused by the adoption of steep blaze angles so we adopt $\tan\gamma = 4$ as a sensible upper limit. Note that the length of the grating along axis is directly proportional to $\tan\gamma$ so this results in a relatively bulky disperser despite the modest beam size. Note also that increasing the dispersion order changes only the ruling density – not the blaze angle – so we show results only for $m_Z = 1$. It should be noted that the results for large $\gamma$ may be invalid because the analysis ignore polarisation effects.

For the IPS, it is assumed that the beam size is 3mm for options B and C. In the tables, $N$ and $N_{min}$ refer to the number of scattering centres if the disperser is filled at the quoted ruling density, $\rho$, and the number required by (12).

For an 8m telescope, the IPS is beneficial for options C and D. For option B, the grating tilt would be large enough to cause potential problems with efficiency and polarisation dependence. For a 40m telescope the utility of an STS is much reduced although marginally feasible for very low spectral resolution.

The IPS appears to be feasible for both 8 and 40m telescopes but clearly advantageous for a 40m telescope where they offer probably the only option. The number of waveguides required to fill the disperser aperture is large but in principle this could be substantially reduced without compromising the resolving power.

Another option available to instrument builders is to use image slicing to reduce the pupil size at the expense of increasing the field parallel to the slit or increasing the number of spectrographs (each of reduced size). This has been discussed at length by Allington-Smith (2007). However this is unlikely to produce the large reductions in size which may be possible with IPS to allow fibre-spectrograph integration..

Table 1: Design parameters for an 8m telescope. Possible photonic spectrograph solutions are highlighted. Various non-photonic solutions are given as described in the text.

|   | $R$ | $\lambda_0$ (nm) | $\chi$ (") | $F'$ | $D$ (mm) | $m$ | $\tan\gamma$ | $q$ ($\mu$m) | $\rho$ (mm$^{-1}$) | $N$ | $N_{min}$ |
|---|---|---|---|---|---|---|---|---|---|---|---|
| A | 50000 | 850.00 | 1.2 | 1.3 | 300.0 | 1 | 3.9 |  | 2278 | 50000 | 50000 |
| B | 5000 | 1700.00 | 0.4 | 3.9 | 100.0 | 1 | 0.4 |  | 425 | 5000 | 5000 |
| C | 1000 | 500.00 | 0.2 | 7.7 | 100.0 | 1 | 0.0 |  | 155 | 1000 | 1000 |
| D | 100 | 700.00 | 1.2 | 1.3 | 50.0 | 1 | 0.0 |  | 133 | 100 | 100 |
| A | 50000 | 850.00 | 1.2 | 1.3 | 1200.0 | 1 | 1.0 |  | 1638 | 50000 | 50000 |
| B | 5000 | 1700.00 | 0.4 | 3.9 | 40.0 | 1 | 1.0 |  | 819 | 5000 | 5000 |
| C | 1000 | 500.00 | 0.2 | 7.7 | 4.0 | 1 | 1.0 |  | 2784 | 1000 | 1000 |
| D | 100 | 700.00 | 1.2 | 1.3 | 2.4 | 1 | 1.0 |  | 1989 | 100 | 100 |
| A | 50000 | 850.00 | 1.2 | 1.3 | 290.0 | 1 | 4.0 |  | 2283 | 50000 | 50000 |
| B | 5000 | 1700.00 | 0.4 | 3.9 | 9.7 | 1 | 4.0 |  | 1141 | 5000 | 5000 |
| C | 1000 | 500.00 | 0.2 | 7.7 | 1.0 | 1 | 4.0 |  | 3880 | 1000 | 1000 |
| D | 100 | 700.00 | 1.2 | 1.3 | 0.6 | 1 | 4.0 |  | 2772 | 100 | 100 |
| A | 50000 | 850.00 | 1.2 | 1.3 | **10.0** | 100 | 85.0 | **55** | 27378 | 500 |  |
| B | 5000 | 1700.00 | 0.4 | 3.9 | **3.0** | 20 | 34.0 | **9** | 2281 | 250 |  |
| C | 1000 | 500.00 | 0.2 | 7.7 | **3.0** | 20 | 10.0 | **16** | 776 | 50 |  |
| D | 100 | 700.00 | 1.2 | 1.3 | **1.0** | 12 | 8.4 | **66** | 554 | 8 |  |

Table 2: As previous table, but for a 40m telescope.

|   | $R$ | $\lambda_0$ (nm) | $\chi$ (") | $F'$ | $D$ (mm) | $m$ | $\tan\gamma$ | $q$ ($\mu$m) | $\rho$ (mm$^{-1}$) | $N$ | $N_{min}$ |
|---|---|---|---|---|---|---|---|---|---|---|---|
| A | 50000 | 850.00 | 1.2 | 0.3 | 300.0 | 1 | 19.4 |  | 2350 | 50000 | 50000 |
| B | 5000 | 1700.00 | 0.4 | 0.8 | 100.0 | 1 | 1.9 |  | 1046 | 5000 | 5000 |
| C | 1000 | 500.00 | 0.2 | 1.5 | 100.0 | 1 | 0.2 |  | 762 | 1000 | 1000 |
| D | 100 | 700.00 | 1.2 | 0.3 | 50.0 | 1 | 0.2 |  | 648 | 100 | 100 |
| A | 50000 | 850.00 | 1.2 | 0.3 | 6000.0 | 1 | 1.0 |  | 1638 | 50000 | 50000 |
| B | 5000 | 1700.00 | 0.4 | 0.8 | 200.0 | 1 | 1.0 |  | 819 | 5000 | 5000 |
| C | 1000 | 500.00 | 0.2 | 1.5 | 20.0 | 1 | 1.0 |  | 2784 | 1000 | 1000 |
| D | 100 | 700.00 | 1.2 | 0.3 | 12.0 | 1 | 1.0 |  | 1989 | 100 | 100 |
| A | 50000 | 850.00 | 1.2 | 0.3 | 1450.0 | 1 | 4.0 |  | 2283 | 50000 | 50000 |
| B | 5000 | 1700.00 | 0.4 | 0.8 | 49.0 | 1 | 4.0 |  | 1141 | 5000 | 5000 |
| C | 1000 | 500.00 | 0.2 | 1.5 | 4.9 | 1 | 4.0 |  | 3878 | 1000 | 1000 |
| D | 100 | 700.00 | 1.2 | 0.3 | 2.9 | 1 | 4.0 |  | 2772 | 100 | 100 |
| A | 50000 | 850.00 | 1.2 | 0.3 | **10.0** | 100 | 85.0 | **274** | 136889 | 500 |  |
| B | 5000 | 1700.00 | 0.4 | 0.8 | **3.0** | 20 | 34.0 | **46** | 11407 | 250 |  |
| C | 1000 | 500.00 | 0.2 | 1.5 | **3.0** | 20 | 10.0 | **78** | 3879 | 50 |  |
| D | 100 | 700.00 | 1.2 | 0.3 | **1.0** | 12 | 8.4 | **332** | 2770 | 8 |  |

How can this be done? $N > N_{min}$ implies that the sample aperture represented by the fibre is larger than the diffraction limit by $N/N_{min}$. In principle, it should be possible to design a set of $N_{min}$ waveguides to act as a sparse disperser, perhaps in a non-redundant baselines configuration, to give the required resolution. This is a common approach in interferometry but must be applied here without wasting light since the reduction in the number of scattering centres implies a low filling factor in the disperser of $N_{min}/N$. This problem might be alleviated by careful design of the waveguide apertures, perhaps by tapering them to fill the disperser surface.

## 5 Conclusions

We confirm the basic principles of an integrated photonic spectrograph by direct simulation using Fresnel propagation and we calculate some possible configurations. These suggest that photonic spectrographs offer a major potential advantage especially for the next generation of extremely large telescopes. However many issues remain to be explored including:

- Detailed design of the photonic disperser including: (a) maximising the fill-factor of the waveguides; (b) design of an efficient sparse disperser; (c) optimisation of the geometry to reduce off-axis aberrations without introducing separate optics.

- Interfacing light from a multimode fibre into a waveguide network that requires single mode propagation for efficient operation. This is related to the night-sky suppression system in which fibre Bragg gratings are inscribed on individual photonic crystal fibres (Bland-Hawthorn et al. 2004) which operate in a single mode (or few modes) extracted from a multimode fibre by a "photonic lantern" transition (Leon-Saval et al. 2005).

- Device fabrication of critical components using techniques such as Ultrafast Laser Inscription (TKA).

- Detectors that can be coupled efficiently to the output of the photonic spectrograph, a problem that is generic to a wide range of future astrophotonic technologies (Le Coärer et al. 2008).

These issues are being addressed by the *Astrophotonica Europa* consortium (Allington-Smith 2008) which has recently been awarded EU FP7 support via the OPTICON network.

## Acknowledgements

We thank many colleagues for useful discussions including Anthony Horton, Robert Thomson and Ajoy Kar. We thank the reviewer for helpful comments.